\documentclass{iau} 
\usepackage{graphicx}
\usepackage{color} 

\usepackage{xspace}
\newcommand{\hii}{H~{\scshape ii}\xspace}
\newcommand{\Ha}{\ifmmode {\mathrm{H}\alpha} \else H$\alpha$\fi\xspace}
\newcommand{\Hb}{\ifmmode {\mathrm{H}\beta} \else H$\beta$\fi\xspace}
\newcommand{\oiii}{\ifmmode [\text{O}\,\textsc{iii}] \else [O~{\scshape iii}]\fi\xspace}
\newcommand{\Oiii}{\ifmmode [\text{O}\,\textsc{iii}]\lambda 5007 \else [O~{\scshape iii}]$\lambda 5007$\fi\xspace}
\newcommand{\nii}{\ifmmode [\text{N}\,\textsc{ii}] \else [N~{\scshape ii}]\fi\xspace}
\newcommand{\Nii}{\ifmmode [\text{N}\,\textsc{ii}]\lambda 6584 \else [N~{\scshape ii}]$\lambda 6584$\fi\xspace}
\newcommand{\Sii}{\ifmmode [\text{S}\,\textsc{ii}]\lambda 6716 \else [S~{\scshape ii}]$\lambda 6716$\fi\xspace}
\newcommand{\WHa}{\ifmmode W_{\mathrm{H}\alpha} \else $W_{\mathrm{H}\alpha}$\fi\xspace}

\title[DIG wrongdoings in galaxies]
{The importance of the diffuse ionized gas for interpreting galaxy spectra}

\author[N.\ Vale Asari \& G.\ Stasi\'nska]
{ Natalia Vale Asari$^{1, 2}$\thanks{Royal Society--Newton Advanced Fellowship}
  \and Gra\.zyna Stasi\'nska$^3$}

\affiliation{$^1$Departamento de F\'isica - CFM - Universidade Federal de Santa Catarina, Florian\'opolis, SC, Brazil \\ email: {\tt natalia@astro.ufsc.br} \\[\affilskip]
  $^2$School of Physics and Astronomy, University of St Andrews, North Haugh, St Andrews KY16 9SS, UK \\[\affilskip]
  $^3$LUTH, Observatoire de Paris, PSL, CNRS 92190 Meudon, France \\ 

}

\pubyear{2020}
\volume{359}
\jname{Galaxy evolution and feedback across different environments}
\editors{T. Storchi-Bergmann, R. Overzier, W. Forman \& R. Riffel, eds.}

\begin{document}

\maketitle

\begin{abstract}

  Diffuse ionized gas (DIG) in galaxies can be found in early-type galaxies, in bulges of late-type galaxies, in the interarm regions of galaxy disks, and outside the plane of such disks. The emission-line spectrum of the DIG can be confused with that of a weakly active galactic nucleus. It can also bias the inference of chemical abundances and star formation rates in star forming galaxies. We discuss how one can detect and feasibly correct for the DIG contribution in galaxy spectra.


\end{abstract}

\firstsection
\section{Introduction}

A lot can be learned from studying the integrated spectra of galaxies.
The power is in the numbers: a large statistical sample tells us about trends in astrophysics, but also about dispersions in those trends.
Empirical relations thus constructed can provide useful guidance for chemical evolution models.

For the sake of the argument let us focus on some empirical laws using the Sloan Digital Sky Survey (SDSS, York et al.\ 2000) data.
One is the stellar mass--nebular metallicity relation (Tremonti et al.\ 2004), which informs us about the history of chemical enrichment of galaxies.
The metallicity, $Z$, depends not only on the yields and on star-formation histories, but also on the inflow and outflow of gas with chemical compositions different from that of the galaxy.

Another important relation is the stellar mass--star formation rate ($M_\star$--SFR) relation (Brinchmann et al.\ 2004). It shows that larger galaxies are also forming more stars. This relation has been later wrongly extrapolated, \Ha being carelessly transformed into SFR to reveal a `quiescent sequence'. This quiescent sequence is nothing more than a sequence of retired galaxies and has nothing to do with star formation (see Section~\ref{sec:retired} below).

One of the most popular empirical relations nowadays is the $M_\star$--$Z$--SFR relation (e.g.\ Mannucci et al.\ 2010). In the representation by Mannucci et al., galaxies in different mass bins show different trends in the $Z$ versus SFR plane.
Low-mass bins show an anticorrelation of $Z$ with SFR, whereas high-mass bins show no trend at all.

As a matter of fact, whereas the concept of a galaxy's total stellar mass and global star formation rate make sense and are intrinsically related to the galaxy as a whole, the `metallicity' is more a fraught term, because the methods to measure the galaxy's `metallicity' have actually been developed  for (giant) \hii regions.
This ignores the fact that the line-emission regions  in a galaxy comprise compact \hii regions, giant \hii regions of diverse morphologies, and diffuse ionized regions.

Several biases may permeate the results in the above and similar papers.
As a whole, one needs to care about sample selection and aperture effects.
For the SFR, one also needs to deal with dust correction (see Vale Asari et al.\ in prep.), with the calibration used for the SFR, and with the contamination by the diffuse ionized gas (DIG).
For the determination of the metallicity, it is well-known that the method, indices and calibration used may change the results dramatically (see e.g. Maiolino \& Mannucci 2019, Kewley et al.\ 2019). So far, the influence of the DIG has not been studied in detail (except by a few like Kumari et al.\ 2019, Poetrodjojo et al.\ 2019, and Vale Asari et al.\ 2019, hereafter VA19) but it could be important.

There is another domain of galaxy research where the DIG is relevant: this is the field of active galactic nuclei (AGN). Weak line emission in the integrated spectra of galaxies has been traditionally interpreted as due to low-level activity linked to accretion onto a supermassive black hole (Kauffmann et al.\ 2003, Kewley et al.\ 2006). However, it has been shown that, in galaxies which have stopped forming stars, dubbed 'retired' galaxies, ionization by hot low-mass evolved stars (HOLMES)  is able to explain both the observed emission line-ratios and their luminosities (Stasi\'nska et al.\ 2008, Cid Fernandes et al.\ 2011).

\section{A condensed history of the DIG}

The DIG was first discovered as a faint extraplanar emission in the Milky Way (Reynolds 1971, 1989) and in edge-on galaxies (Dettmar 1990, Hoopes et al.\ 1996, 1999).
In the context of our Galaxy it is often referred to as the warm ionized medium or diffuse ionized medium.
It has later been found in interarm regions, where the \hii regions do not outshine it, or the density of the gas is smaller (Walterbros \& Brown 1994, Wang et al.\ 1999, Zurita et al.\ 2000).
Some studies find that 30 to 60 per cent of the total \Ha in emission in a spiral galaxy may be due to the DIG (e.g. Oey et al.\ 2007).
Warm ionized gas has also been detected in early-type galaxies (Phillips et al.\ 1986, Martel et al.\ 2004, Jaff{\'e} et al.\ 2014, Johansson et al.\ 2016).

The recent development of integral field spectroscopy (IFS) boosted studies of resolved properties of nearby galaxies in which the properties of bona fide \hii regions and DIG can be separated (e.g.\ Blanc et al.\ 2009, Kaplan et al.\ 2016; Kreckel et al.\ 2016; Poetrodjojo et al.\ 2019).

Already several decades ago it was found that the DIG has a lower electron density, higher electron temperature, and enhanced collisionally-excited to recombination emission line ratios (\Nii/\Ha, \Sii\Ha, also usually \Oiii/\Hb) as compared to \hii regions (Galarza et al.\ 1998).
This suggests that the DIG is ionized by a mechanism other than photoionization by  OB stars. Propositions include 
cosmic rays (Reynolds \& Cox 1992, Vandenbroucke et al.\ 2018), photoionization by old supernova remnants (Slavin et al.\ 2000), 
dissipation of turbulence (Minter \& Spangler 1997, Minter \& Balser 1997, Binette et al.\ 2009), contribution of dust-scattered light (Wood \& Reynolds 1999), 
shocks from supernova winds (Collins \& Rand 2001),
ionization by photons leaking from star-forming (SF) regions (Domgorgen \& Mathis 1994, Haffner et al.\ 2009, Weilbacher et al.\ 2018), and photoionization by HOLMES (Binette et al.\ 1994, Stasi\'nska et al.\  2008, Athey \& Bregman 2009, Flores-Fajardo et al.\ 2011, Yan \& Blanton 2013).

\section{The DIG in early-type galaxies and in bulges}
\label{sec:retired}

\begin{figure}[tb]
  \begin{center}
    \includegraphics[width=0.8\textwidth]{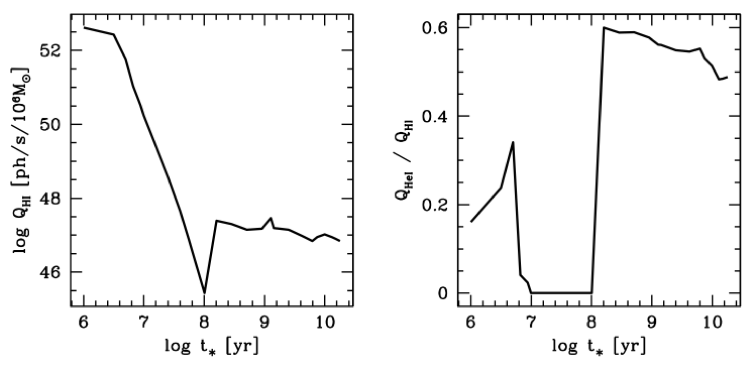} 
    \caption{\textit{Left:} The rate $Q_{\mathrm H}$ of  photons capable of ionizing a hydrogen atom  for a simple stellar population from Bruzual \& Charlot (2003). \textit{Right: }The hardness $Q_{\mathrm He}/Q_{\mathrm H}$ of the ionizing radiation field as a function of age for the same SSP.}
    \label{QH}
  \end{center}
\end{figure}

Emission-line ratios can serve to distinguish the main ionization mechanism in galaxies. The most famous  diagram,  \Nii/\Ha versus \Oiii/\Hb (Baldwin et al.\ 1981; BPT)  drew an empirical line to separate giant \hii regions from planetary nebulae and objects ionized by a power-law spectrum or excited by shocks.

With the advent of the SDSS and its wealth of spectroscopic data, the separation of the BPT plane in several zones became much clearer (Kauffmann et al.\ 2003, Kewley et al.\ 2006).
It has been commonly said that SF galaxies lie  in the same region as giant \hii regions, and objects on the right-hand side of the diagram are AGNs, subdivided into Seyfert and LINERs. Note that the acronym LINER stands for low-ionization \textit{nuclear} emission regions (Heckman 1981), and a priori does not apply to SDSS spectral observations, which where made through 3-arcsec fibers and covered a significant portion of the galaxies -- except for the nearest ones. 
So the part of the diagram where SDSS galaxies with LINER-like spectra lie cannot all be populated by \textit{bona fide} LINERs.

Stasi\'nska et al.\ (2008)  proposed that HOLMES\footnote{Stasi\'nska et al.\ (2008) did not use the HOLMES terminology at that time, but the expression `post-AGB stars', which has a double meaning in astrophysics.} could be responsible for the observed LINER-like emission-line ratios.
Fig.~\ref{QH} shows the reasoning behind it. The panel on the left shows the rate $Q_{\mathrm H}$ of  photons capable of ionizing a hydrogen atom as a function of time, for a simple stellar population (SSP) from Bruzual \& Charlot (2003, BC03). Here  a Chabrier (2003) initial-mass function (IMF)  and solar metallicity are used. $Q_{\mathrm H}$ is seen to fall by 5 orders of magnitude between  $10$~Myr and $100$~Myr. However, there is a continuous production of ionizing photons for ages larger than $100$~Myr. These arise from stars that have evolved off the asymptotic giant branch, have become very hot (some of them may reach 200,000 K) and are on the way of becoming degenerate stars of freshly-formed white dwarfs. Such stars are faint in comparison to OB stars, but their huge numbers from the IMF  make up for their faintness.

More importantly, the right panel of Fig.~\ref{QH} shows the hardness of the ionizing radiation field as a function of age for the same SSP, where $Q_\mathrm{He}$ is the rate of photons with energies $> 24.6$~eV.
The ionizing photons from HOLMES are on average more energetic than those from young stars, which means that on average the electron kinetic energy will be greater in a gas ionized by HOLMES than by OB stars. This implies that the collisionally-excited lines emitted by this gas will be stronger with respect to recombination lines than in the case of \hii regions. 

\begin{figure}[tb]
\begin{center}
  \includegraphics[width=0.4\textwidth]{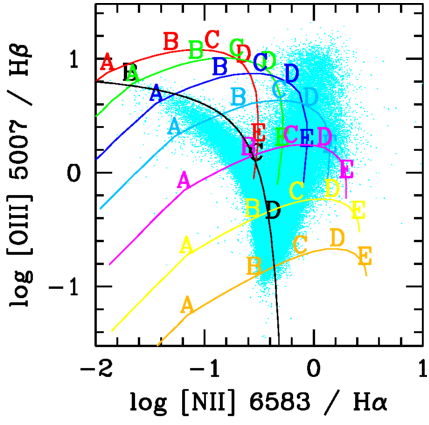} 
  \includegraphics[width=0.42\textwidth]{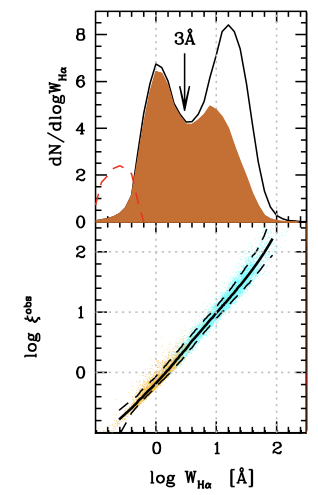} 
  \caption{\textit{Left}: Sequences of photoionization models (of given metallicity and varying ionization parameter) where the ionization source are HOLMES in the BPT plane plotted over observed SDSS galaxies (Figure from Stasi\'nska et al.\ 2008).
  \textit{Right}: Top panel shows the \WHa histogram for SDSS galaxies. The bottom panel shows $\xi$ (the ratio of the total observed \Ha luminosity to the expected \Ha luminosity due to ionization by HOLMES)  versus \WHa for the same galaxies, where the solid (dashed) line is the median (10 and 90 percentile) relation.
  The $3$~\AA\ arrow delimits the separation between galaxies solely explained by ionization by HOLMES from galaxies where an extra ionization source is needed.
  (Figure adapted from Cid Fernandes et al.\ 2011.)
}
  \label{BPT-S08}
\end{center}
\end{figure}

The first \textit{ab initio} models for retired galaxies ionized by HOLMES were made by Stasi\'nska et al (2008). Fig.~\ref{BPT-S08} (left) shows an example of line ratios calculated from photoionization models using as an ionization source the spectrum obtained by the spectral synthesis code {\sc starlight} (Cid Fernandes et al.\ 2005) for galaxies in the LINER-like region of the BPT. The stellar populations were modeled in the optical region to reproduce the SDSS spectra, and the ionizing part of the spectrum was extrapolated from the BC03 stellar population models.
The photoionization models show that radiation from HOLMES can explain the whole BPT plane, except for the rightmost tip where Seyfert galaxies live.

Fig.~\ref{BPT-S08} (right), from Cid Fernandes et al.\ (2011), shows that not only line ratios for galaxies with LINER-like spectra can be explained only by HOLMES, but also the budget of their ionizing photons. The parameter $\xi$ is the total observed \Ha luminosity divided by the expected \Ha luminosity assuming that all the photons from HOLMES ($> 100$~Myr) are absorbed by the gas. This parameter $\xi$ is actually very well correlated with the \Ha equivalent width (\WHa). The sample shown in this figure is a mixture of all SDSS galaxies. The bimodality shown in the \WHa distribution separates galaxies where HOLMES can be the sole ionization mechanism (below 3 \AA), and other galaxies where extra sources are needed to explain the enhanced \Ha luminosity (SF, AGN, etc.).

\section{The DIG in face-on late-type galaxies}

Many of the problems found in the SDSS studies have come about because we are unable to separate the ionizing sources in a galaxy. Several studies based on IFS (e.g.\ Sarzi et al. 2010; Belfiore et al.\ 2016; Gomes et al.\ 2016) have found evidence of real LINERS in the nuclei of some galaxies, and of LINER-like emission \textit{outside} the nuclei of late-type galaxies (dubbed `LIER', where the \emph{N} for \emph{nuclear} has been dropped).
Recent IFS studies of SF galaxies have found that the DIG has kinematic properties which are different from those of \hii regions, having been found to come from a thicker layer (den Brok et al.\ 2020) and to sustain more turbulence (Della Bruna et al.\ 2020).
In the following we show how DIG biases metallicity measurements in SF galaxies based on emission line ratios, and present ways to mitigate those biases using IFS data.

\subsection{Digging out the DIG with integral field spectroscopy}

There are several ways to identify regions where the DIG emission is important.
One of them is based on line ratios, such as \Sii/\Ha (e.g.\ Kreckel et al.\ 2016). However, if we would ultimately like to quantify the contribution of the DIG to line ratios, tagging DIG regions using this criterion would make our analysis circular.

A second method uses the \Ha surface brightness (e.g.\ Zhang et al.\ 2017). This criterion may break down due to a simple geometrical effect, e.g.\ in the bulge of late-type galaxies the column density of the gas in the line-of-sight is larger, which makes \Ha brighter. Therefore, diffuse emission from bulges would be missed with a simple \Ha surface brightness cut.

\begin{figure}[b]
  \begin{center}
    \includegraphics[width=0.6\textwidth]{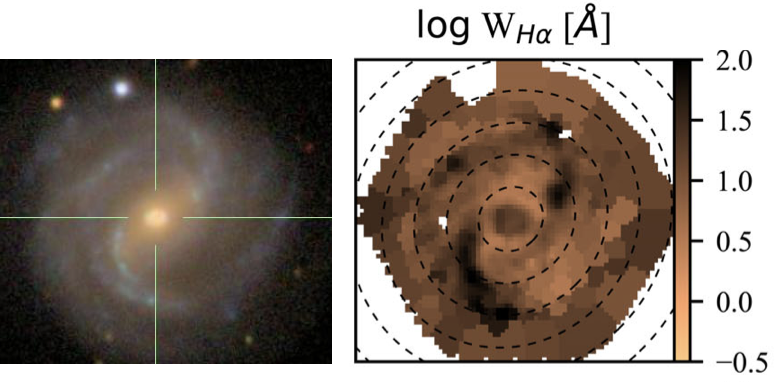} 
    \caption{SDSS image and \WHa map for the CALIFA galaxy 0073. (Figure adapted from Lacerda et al.\ 2018.)}
    \label{WHa-map-L18}
  \end{center}
\end{figure}

A third criterion, based on \WHa, is the one we favour. Fig.~\ref{WHa-map-L18} shows a SDSS colour-image and a \WHa map from the Calar Alto Legacy Integral Field Area (CALIFA, S{\'a}nchez, et al.\ 2016) survey for the same galaxy; one may note that high \WHa regions trace the spiral arms quite well.
Lacerda et al. (2018) have thus proposed a classification built upon of the bimodality of the \WHa distribution, which is found for both SDSS spectra and CALIFA spaxels.  CALIFA spaxels have been sorted into three classes: (1) HOLMES DIG (hDIG) where $\WHa < 3$~\AA, (2) mixed DIG (mDIG) where $3 < \WHa < 14$~\AA, (3) SF complexes (SFc) where $\WHa > 14$~\AA.

Spaxel sizes in CALIFA are $\sim 1$-kpc wide, so hDIG and mDIG regions may still contain buried-in \hii regions. SFc spaxels, on the other hand, are not classical \hii regions (which usually have $\WHa \sim 100$--1000~\AA), but may encompass many \hii regions and some DIG.

\begin{figure}[tb]
  \begin{center}
    \includegraphics[width=0.5\textwidth]{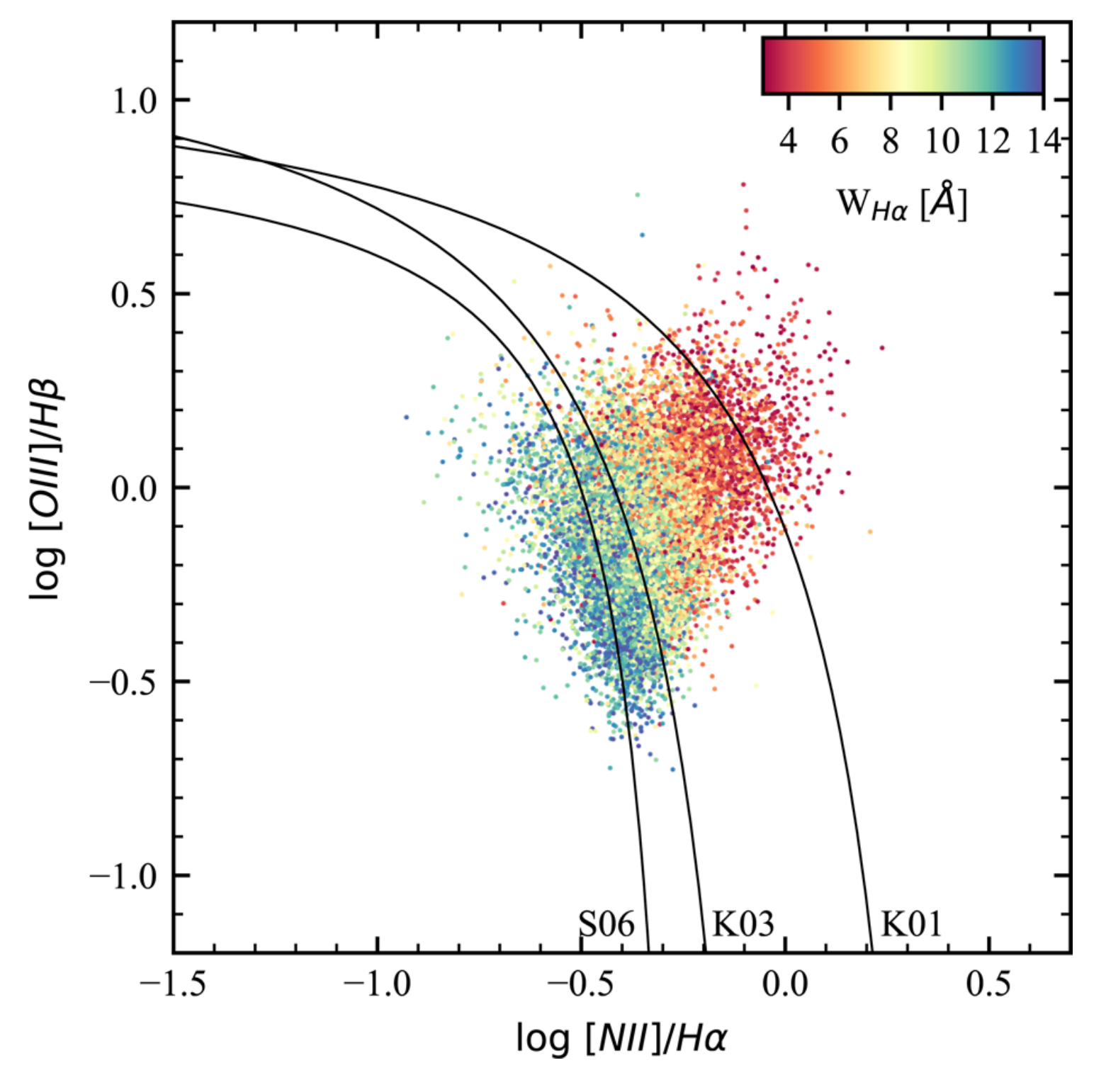} 
    \caption{BPT diagram for CALIFA regions with \WHa in the $3–14$~\AA\ range, coloured according to \WHa, and excluding zones inwards of one half light radius. (Figure from Lacerda et al.\ 2018.)}
    \label{BPT-L18}
  \end{center}
\end{figure}

Fig.~\ref{BPT-L18} from Lacerda et al.\ (2018) shows only spaxels classified as mDIG for CALIFA galaxies on the BPT plane, colour-coded by \WHa. Smaller \WHa values (i.e.\ where larger DIG contribution is expected) correspond to larger \nii/\Ha and \oiii/\Hb\ line ratios.
This means that spaxels with more DIG emission do have systematically different emission line ratios, which must bias studies using those ratios as proxies for gas-phase metallicity.

\subsection{A method to remove the DIG contribution in the integrated spectrum of a galaxy}

To be able to extract the contribution of \textit{bona fide} \hii regions from an observed emission-line spectrum, one needs an empirical method. VA19 developed a method using Mapping Nearby Galaxies at APO (MaNGA, Blanton et al. 2017) IFS data. Similarly to CALIFA these data have $\sim 1$~kpc resolution and the contamination of the DIG to SFc spaxels still holds true, the advantage of using MaNGA being essentially a larger sample of galaxies.

\begin{figure}[b]
\begin{center}
  \includegraphics[width=0.6\textwidth]{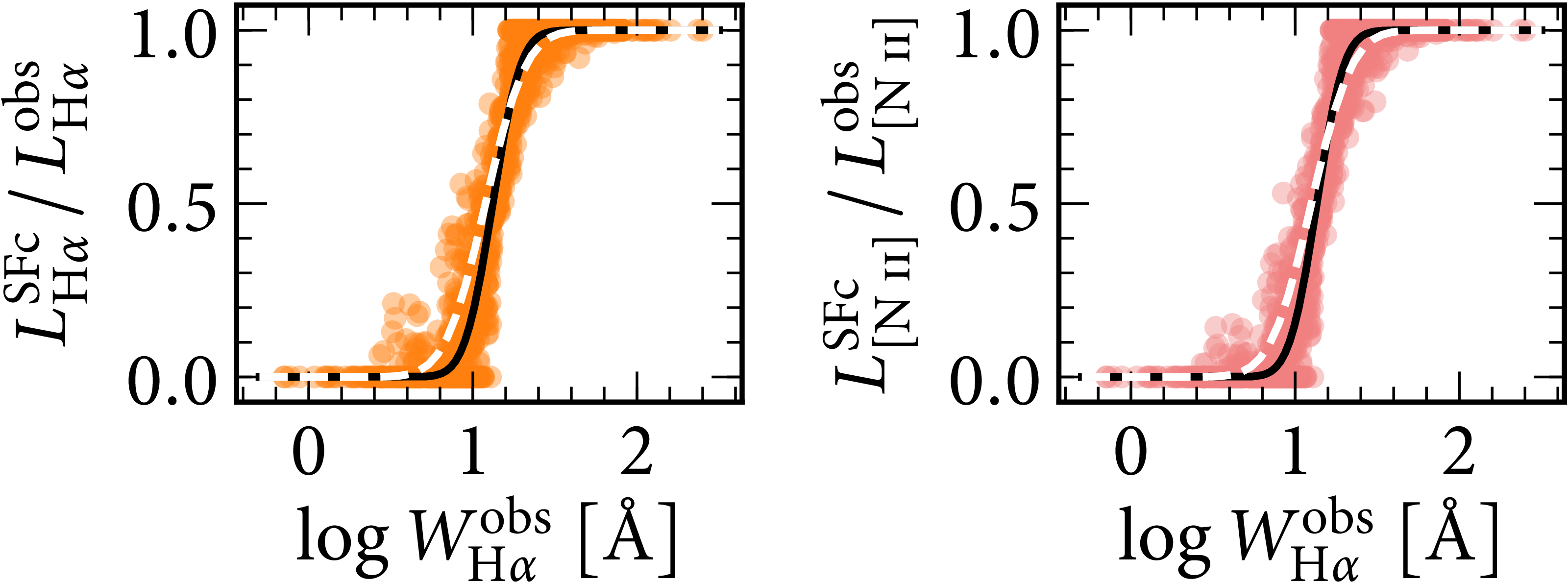}
  \caption{
  Correction for DIG emission appropriate for SDSS observations, based on 1409 MaNGA star-forming galaxies. Panels show the \Ha and \Nii line luminosities in SFc spaxels normalised by the total luminosity versus total \WHa, where all measurements were taken within a circular $0.7 R_{50}$-diameter aperture.
  The solid line shows a fit to the data; the dashed line shows a fit for measurements made in $2.0 R_{50}$-diameter apertures.
  (Figure adapted from VA19.)
}
  \label{DIGcor-V19}
\end{center}
\end{figure}

Fig.~\ref{DIGcor-V19} shows the correction proposed for \Nii and \Ha emission lines (other lines are in VA19). The abscissa for both panels is the global observed \WHa, i.e.\ measured in circular $0.7 R_{50}$-diameter apertures. 
The ordinate shows the ratio $L_\mathrm{SFc}/L_\mathrm{obs}$, where $L_\mathrm{obs}$ is the total luminosity in a line, and $L_\mathrm{SFc}$ is the line luminosity adding up only spaxels tagged as SFc (i.e.\ removing hDIG and mDIG spaxels). The left panel concerns the \Ha line and the right one \nii.
This ratio increases from zero -- where there is no contribution from SFc to the total spectra -- to one -- the whole emission is in SFc spaxels. Crucially, these curves are slightly different for each emission line.

\begin{figure}[htb]
\begin{center}
  \includegraphics[width=1\textwidth, trim=50 0 220 0, clip]{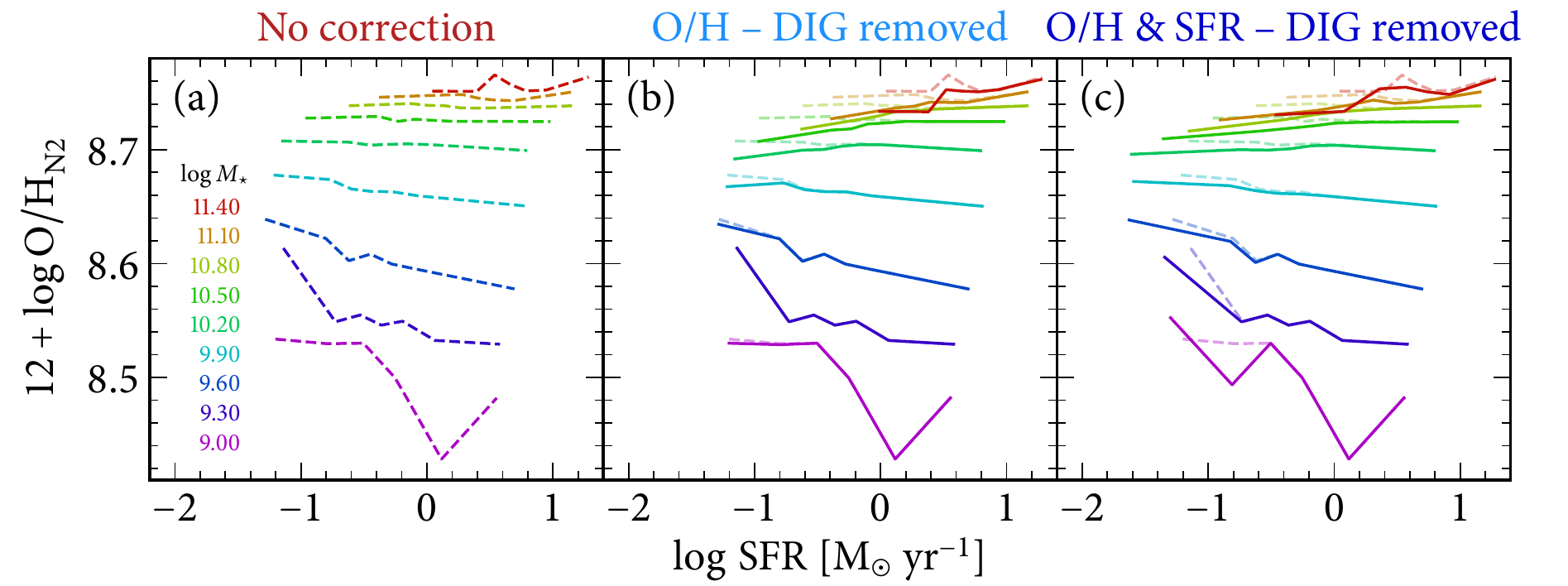}
  \caption{O/H as a function of SFR for $\sim 10,000$ SDSS star-forming galaxies in stellar mass bins (whose centres are given in the left handside panel). O/H has been calculated using the \Nii/\Ha index.
  (a) $M$--$Z$--SFR relation with no correction. This is repeated (as dashed translucent lines) in the two other panels for comparison.
  (b) Correcting \Nii and \Ha for the DIG contamination (using the fits shown in Fig.~\ref{DIGcor-V19}) prior to calculating O/H.
  (c) Correcting \Ha prior to obtaining the SFR as well.
  (Figure adapted from VA19.)
}
  \label{MZSFR-V19}
\end{center}
\end{figure}

Fig.~\ref{MZSFR-V19} shows the $M$--$Z$--SFR relation for SDSS galaxies, where the oxygen abundance is calculated using the \nii/\Ha line ratio.
Panel (a) shows the uncorrected relation, which is repeated in translucent dashed lines in the other panels.
Panel (b) overplots, in solid lines, the relation where O/H has been recalculated by removing the contribution from the DIG to \nii\ and \Ha using the fit from Fig.~\ref{DIGcor-V19}.
High-mass bins are the most affected, now featuring a correlation which was absent in panel (a).
Panel (c) shows the effect of also removing the DIG contribution to \Ha prior to computing the SFR.

Changes to the $M$--$Z$--SFR are small because, as mentioned before, for MaNGA observations even SFc spaxels still contain non-negligible contribution from the DIG. A similar approach using a large sample of data obtained with the Multi Unit Spectroscopic Explorer (MUSE, Bacon et al.\ 2010) should give a 
larger difference and allow a more reliable correction.
Note that this method is purely empirical and does not rely on any assumption regarding the source of ionization of the DIG.

{\small {\bf Acknowledgements:} NVA acknowledges support of FAPESC and CNPq, and of the Royal Society--Newton Advanced Fellowship award (NAF\textbackslash{}R1\textbackslash{}180403). GS acknowledges a CNPq visiting professor grant.}


\begin{thebibliography}{}

\bibitem[\protect\citeauthoryear{Athey \& Bregman}{2009}]{2009ApJ...696..681A} Athey A.~E., Bregman J.~N., 2009, ApJ, 696, 681

\bibitem[\protect\citeauthoryear{Bacon, et al.}{2010}]{2010SPIE.7735E..08B} Bacon R., et al., 2010, SPIE, 7735, 773508, SPIE.7735
  
\bibitem[Baldwin et al.(1981)]{1981PASP...93....5B} Baldwin, J.~A., Phillips, M.~M., \& Terlevich, R.\ 1981, PASP, 93, 5

\bibitem[\protect\citeauthoryear{Belfiore, et al.}{2016}]{2016MNRAS.461.3111B} Belfiore F., et al., 2016, MNRAS, 461, 3111

\bibitem[\protect\citeauthoryear{Blanc, et al.}{2009}]{2009ApJ...704..842B} Blanc G.~A., Heiderman A., Ge
  bhardt K., Evans N.~J., Adams J., 2009, ApJ, 704, 842

  \bibitem[\protect\citeauthoryear{Blanton, et al.}{2017}]{2017AJ....154...28B} Blanton M.~R., et al., 2017, AJ, 154, 28
  
\bibitem[\protect\citeauthoryear{Binette, et al.}{1994}]{1994A&A...292...13B} Binette L., Magris C.~G., Stasi{\'n}ska G., Bruzual A.~G., 1994, A\&A, 292, 13

\bibitem[Binette et al.(2009)]{2009ApJ...695..552B} Binette, L., Flores-Fajardo, N., Raga, A.~C., et al.\ 2009, ApJ, 695, 552

\bibitem[\protect\citeauthoryear{Brinchmann, et al.}{2004}]{2004MNRAS.351.1151B} Brinchmann J., Charlot S., White S.~D.~M., Tremonti C., Kauffmann G., Heckman T., Brinkmann J., 2004, MNRAS, 351, 1151
  
\bibitem[\protect\citeauthoryear{Bruzual \& Charlot}{2003}]{2003MNRAS.344.1000B} Bruzual G., Charlot S., 2003, MNRAS, 344, 1000

\bibitem[\protect\citeauthoryear{Chabrier}{2003}]{2003PASP..115..763C} Chabrier G., 2003, PASP, 115, 763

\bibitem[\protect\citeauthoryear{Cid Fernandes, et al.}{2005}]{2005MNRAS.358..363C} Cid Fernandes R., Mateus A., Sodr{\'e} L., Stasi{\'n}ska G., Gomes J.~M., 2005, MNRAS, 358, 363

\bibitem[\protect\citeauthoryear{Cid Fernandes, et al.}{2011}]{2011MNRAS.413.1687C} Cid Fernandes R., Stasi{\'n}ska G., Mateus A., Vale Asari N., 2011, MNRAS, 413, 1687

\bibitem[Collins \& Rand(2001)]{2001ApJ...551...57C} Collins, J.~A., \& Rand, R.~J.\ 2001, ApJ, 551, 57

\bibitem[\protect\citeauthoryear{Della Bruna, et al.}{2020}]{2020AA...635A.134D} Della Bruna L., et al., 2020, A\&A, 635, A134

\bibitem[\protect\citeauthoryear{den Brok, et al.}{2020}]{2020MNRAS.491.4089D} den Brok M., et al., 2020, MNRAS, 491, 4089

\bibitem[\protect\citeauthoryear{Dettmar}{1990}]{1990AA...232L..15D} Dettmar R.-J., 1990, A\&A, 232, L15

\bibitem[\protect\citeauthoryear{Domgorgen \& Mathis}{1994}]{1994ApJ...428..647D} Domgorgen H., Mathis J.~S., 1994, ApJ, 428, 647

\bibitem[Dopita et al.(2013)]{2013ApJS..208...10D} Dopita, M.~A., Sutherland, R.~S., Nicholls, D.~C., et al.\ 2013, ApJS, 208, 10

\bibitem[Flores-Fajardo et al.(2011)]{2011MNRAS.415.2182F} Flores-Fajardo, N., Morisset, C., Stasi{\'n}ska, G., et al.\ 2011, MNRAS, 415, 2182

\bibitem[\protect\citeauthoryear{Galarza, Walterbos \& Braun}{1998}]{1998AAS...192.4007G} Galarza V.~C., Walterbos R.~A.~M., Braun R., 1998, A\&AS, 192, 40.07

\bibitem[\protect\citeauthoryear{Gomes, et al.}{2016}]{2016A&A...588A..68G} Gomes J.~M., et al., 2016, A\&A, 588, A68
  
\bibitem[Haffner et al.(2009)]{2009RvMP...81..969H} Haffner, L.~M., Dettmar, R.-J., Beckman, J.~E., et al.\ 2009, Reviews of Modern Physics, 81, 969

\bibitem[Heckman(1980)]{1980A&A....87..152H} Heckman, T.~M.\ 1980, A\&A, 500, 187

\bibitem[Hoopes et al.(1999)]{1999ApJ...522..669H} Hoopes, C.~G., Walterbos, R.~A.~M., \& Rand, R.~J.\ 1999, ApJ, 522, 669
\bibitem[Hoopes et al.(1996)]{1996AJ....112.1429H} Hoopes, C.~G., Walterbos, R.~A.~M., \& Greenwalt, B.~E.\ 1996, AJ, 112, 1429

\bibitem[\protect\citeauthoryear{Jaff{\'e}, et al.}{2014}]{2014MNRAS.440.3491J} Jaff{\'e} Y.~L., et al., 2014, MNRAS, 440, 3491

\bibitem[\protect\citeauthoryear{Johansson, et al.}{2016}]{2016MNRAS.461.4505J} Johansson J., Woods T.~E., Gilfanov M., Sarzi M., Chen Y.-M., Oh K., 2016, MNRAS, 461, 4505

\bibitem[\protect\citeauthoryear{Kaplan, et al.}{2016}]{2016MNRAS.462.1642K} Kaplan K.~F., et al., 2016, MNRAS, 462, 1642
  
\bibitem[Kauffmann et al.(2003)]{2003MNRAS.346.1055K} Kauffmann, G., Heckman, T.~M., Tremonti, C., et al.\ 2003, MNRAS, 346, 1055

\bibitem[Kewley et al.(2001)]{2001ApJ...556..121K} Kewley, L.~J., Dopita, M.~A., Sutherland, R.~S., et al.\ 2001, ApJ, 556, 121

\bibitem[\protect\citeauthoryear{Kewley, et al.}{2006}]{2006MNRAS.372..961K} Kewley L.~J., Groves B., Kauffmann G., Heckman T., 2006, MNRAS, 372, 961

\bibitem[\protect\citeauthoryear{Kewley, Nicholls \& Sutherland}{2019}]{2019ARA&A..57..511K} Kewley L.~J., Nicholls D.~C., Sutherland R.~S., 2019, ARAA, 57, 511

\bibitem[\protect\citeauthoryear{Kreckel, et al.}{2016}]{2016ApJ...827..103K} Kreckel K., Blanc G.~A., Schinnerer E., Groves B., Adamo A., Hughes A., Meidt S., 2016, ApJ, 827, 103

\bibitem[\protect\citeauthoryear{Kumari, et al.}{2019}]{2019MNRAS.485..367K} Kumari N., Maiolino R., Belfiore F., Curti M., 2019, MNRAS, 485, 367

\bibitem[\protect\citeauthoryear{Lacerda, et al.}{2018}]{2018MNRAS.474.3727L} Lacerda E.~A.~D., et al., 2018, MNRAS, 474, 3727
  
\bibitem[\protect\citeauthoryear{Maiolino  Mannucci}{2019}]{2019AARv..27....3M} Maiolino R., Mannucci F., 2019, AARv, 27, 3

\bibitem[\protect\citeauthoryear{Mannucci, et al.}{2010}]{2010MNRAS.408.2115M} Mannucci F., Cresci G., Maiolino R., Marconi A., Gnerucci A., 2010, MNRAS, 408, 2115


\bibitem[\protect\citeauthoryear{Martel, et al.}{2004}]{2004AJ....128.2758M} Martel A.~R., et al., 2004, AJ, 128, 2758

\bibitem[\protect\citeauthoryear{Minter \& Balser}{1997}]{1997ApJ...484L.133M} Minter A.~H., Balser D.~S., 1997, ApJL, 484, L133


\bibitem[\protect\citeauthoryear{Minter \& Spangler}{1997}]{1997ApJ...485..182M} Minter A.~H., Spangler S.~R., 1997, ApJ, 485, 182

\bibitem[Oey et al.(2007)]{2007ApJ...661..801O} Oey, M.~S., Meurer, G.~R., Yelda, S., et al.\ 2007, ApJ, 661, 801

\bibitem[\protect\citeauthoryear{Phillips, et al.}{1986}]{1986AJ.....91.1062P} Phillips M.~M., Jenkins C.~R., Dopita M.~A., Sadler E.~M., Binette L., 1986, AJ, 91, 1062

\bibitem[\protect\citeauthoryear{Poetrodjojo, et al.}{2019}]{2019MNRAS.487...79P} Poetrodjojo H., et al., 2019, MNRAS, 487, 79

\bibitem[Reynolds(1971)]{1971PhDT.........1R} Reynolds, R.~J.\ 1971, Ph.D. Thesis

\bibitem[Reynolds \& Cox(1992)]{1992ApJ...400L..33R} Reynolds, R.~J., \& Cox, D.~P.\ 1992, ApJl, 400, L33

\bibitem[\protect\citeauthoryear{S{\'a}nchez, et al.}{2016}]{2016A&A...594A..36S} S{\'a}nchez S.~F., et al., 2016, A\&A, 594, A36
  
\bibitem[\protect\citeauthoryear{Sarzi, et al.}{2010}]{2010MNRAS.402.2187S} Sarzi M., et al., 2010, MNRAS, 402, 2187
  
\bibitem[\protect\citeauthoryear{Slavin, McKee \& Hollenbach}{2000}]{2000ApJ...541..218S} Slavin J.~D., McKee C.~F., Hollenbach D.~J., 2000, ApJ, 541, 218

\bibitem[Stasi{\'n}ska et al.(2006)]{2006MNRAS.371..972S} Stasi{\'n}ska, G., Cid Fernandes, R., Mateus, A., et al.\ 2006, MNRAS, 371, 972

\bibitem[Stasi{\'n}ska et al.(2008)]{2008MNRAS.391L..29S} Stasi{\'n}ska, G., Vale Asari, N., Cid Fernandes, R., et al.\ 2008, MNRAS, 391, L29

\bibitem[\protect\citeauthoryear{Tremonti, et al.}{2004}]{2004ApJ...613..898T} Tremonti C.~A., et al., 2004, ApJ, 613, 898

\bibitem[\protect\citeauthoryear{Vale Asari, et al.}{2019}]{2019MNRAS.489.4721V} Vale Asari N., et al., 2019, MNRAS, 489, 4721 (VA19)

\bibitem[\protect\citeauthoryear{Vandenbroucke, et al.}{2018}]{2018MNRAS.476.4032V} Vandenbroucke B., Wood K., Girichidis P., Hill A.~S., Peters T., 2018, MNRAS, 476, 4032

\bibitem[\protect\citeauthoryear{Walterbos \& Braun}{1994}]{1994ApJ...431..156W} Walterbos R.~A.~M., Braun R., 1994, ApJ, 431, 156

\bibitem[\protect\citeauthoryear{Wang, Heckman \& Lehnert}{1999}]{1999ApJ...515...97W} Wang J., Heckman T.~M., Lehnert M.~D., 1999, ApJ, 515, 97

\bibitem[Weilbacher et al.(2018)]{2018A&A...611A..95W} Weilbacher, P.~M., Monreal-Ibero, A., Verhamme, A., et al.\ 2018, A\&A, 611, A95

\bibitem[\protect\citeauthoryear{Wood \& Reynolds}{1999}]{1999ApJ...525..799W} Wood K., Reynolds R.~J., 1999, ApJ, 525, 799

\bibitem[\protect\citeauthoryear{Yan \& Blanton}{2013}]{2013IAUS..295..328Y} Yan R., Blanton M.~R., 2013, IAUS, 295, 328, IAUS..295

\bibitem[\protect\citeauthoryear{York, et al.}{2000}]{2000AJ....120.1579Y} York D.~G., et al., 2000, AJ, 120, 1579
 
\bibitem[\protect\citeauthoryear{Zhang, et al.}{2017}]{2017MNRAS.466.3217Z} Zhang K., et al., 2017, MNRAS, 466, 3217 

\bibitem[\protect\citeauthoryear{Zurita, Rozas & Beckman}{2000}]{2000A&A...363....9Z} Zurita A., Rozas M., Beckman J.~E., 2000, A\&A, 363, 9
 
\end{thebibliography}
\end{document}